\documentclass[aps,prl,reprint,nofootinbib,preprintnumbers,floatfix]{revtex4-1}

\usepackage[bookmarks=false]{hyperref}
\usepackage{graphicx}
\usepackage{amsmath}
\usepackage{xspace}
\usepackage{color}
\usepackage[small]{subfigure}

\newcommand{\eq}[1]{Eq.~\eqref{eq:#1}}

\newcommand{\fig}[1]{Fig.~\ref{fig:#1}}

\newcommand{\cP}{{\mathcal P}}

\newcommand{\nn}{\nonumber}

\definecolor{purple}{rgb}{0.5,0,0.5}
\definecolor{red}{rgb}{1,0,0}

\allowdisplaybreaks[4]

\newcommand{\fd}[2]{\parbox{#1}{\includegraphics[width=#1]{#2}}}

\newcommand{\ecf}[2]{e_{#1}^{(#2)}} 
\newcommand{\ecfnobeta}[1]{e_{#1}}

\newcommand{\Dobsnobeta}[1]{D_{#1}} 
\newcommand{\ecfop}[1]{\hat{\mathbf{E}}_{#1}}

\newcommand{\eventtwo}{\texttt{EVENT2}}

\def\zcut{z_{\text{cut}}}

\newcommand{\pythia}[1]{\textsc{Pythia\xspace #1}}

\begin{document}


\title{Analytic Boosted Boson Discrimination at the Large Hadron Collider \vspace{0.3cm}}

\author{Andrew J. Larkoski}
\email{larkoski@reed.edu}
\author{Ian Moult}
\email{ianmoult@lbl.gov}
\author{Duff Neill}
\email{duff.neill@gmail.com}
\affiliation{Physics Department, Reed College, Portland, OR 97202,USA\vspace{0.5ex}}
\affiliation{Berkeley Center for Theoretical Physics, University of California, Berkeley, CA 94720, USA\vspace{0.5ex}}
\affiliation{Theoretical Physics Group, Lawrence Berkeley National Laboratory, Berkeley, CA 94720, USA\vspace{0.5ex}}
\affiliation{Theoretical Division, MS B283, Los Alamos National Laboratory, Los Alamos, NM 87545, USA\vspace{0.5ex}}


\begin{abstract}

Jet substructure is playing a central role at the Large Hadron Collider (LHC) probing the Standard Model in extreme regions of phase space and providing innovative ways to search for new physics. Analytic calculations of experimentally successful observables are a primary catalyst driving developments in jet substructure, allowing for a deeper understanding of observables and for the exploitation of increasingly subtle features of jets. In this paper we present a field theoretic framework enabling systematically improvable calculations of groomed multi-prong substructure observables, which builds on recent developments in multi-scale effective theories. We use this framework to compute for the first time the full spectrum for groomed tagging observables at the LHC, carefully treating both perturbative and non-perturbative contributions in all regions. Our analysis enables a precision understanding which we hope will improve the reach and sophistication of jet substructure techniques at the LHC.

\end{abstract}

\maketitle

\newpage

\section{Introduction}
\label{sec:intro}

The Large Hadron Collider (LHC) represents a unique opportunity to probe the detailed structure of the Standard Model at the TeV scale. Collisions at the LHC are dominated by Quantum Chromodynamics (QCD), and in particular, jets, whose radiation encodes the details of the underlying scattering process. There has therefore been a significant theoretical effort to understand more complicated final states involving jets, both in terms of calculating the underlying ultraviolet process \cite{Berger:2010zx,Bern:2011ep,Bern:2013gka,Badger:2013yda,Boughezal:2015ded,Boughezal:2015dva,Boughezal:2015aha,Boughezal:2015dra,Campbell:2016lzl,Currie:2017eqf} and disentangling infrared divergences therein \cite{GehrmannDeRidder:2005cm,Czakon:2010td,Boughezal:2011jf,Czakon:2014oma,Boughezal:2015aha,Gaunt:2015pea,Moult:2016fqy,Boughezal:2016zws,DelDuca:2016ily,Caola:2017dug}, as well as developing new tools for understanding factorization and the infrared dynamics of QCD radiation \cite{Bauer:2000yr,Bauer:2001ct,Bauer:2001yt,Bauer:2002nz,Rothstein:2016bsq,Bauer:2011uc,Larkoski:2015zka,Larkoski:2015kga,Pietrulewicz:2016nwo,Dasgupta:2014yra,Banfi:2014sua,Chien:2015cka,Caron-Huot:2015bja,Larkoski:2015zka,Becher:2015hka,Becher:2016mmh,Chang:2013iba,Larkoski:2016zzc,Larkoski:2015lea}. This has enabled realistic first principles calculations of physical observables on jets \cite{Marzani:2017mva,Frye:2016aiz,Frye:2016okc,Banfi:2016zlc,Banfi:2015pju,Stewart:2013faa,Becher:2012qa,Feige:2012vc,Dasgupta:2013ihk,Dasgupta:2015lxh,Larkoski:2015kga,Frye:2017yrw,Jouttenus:2013hs,Hoang:2017kmk}. 

Experimental advances at the LHC have allowed the detailed substructure of a jet to be measured and exploited to determine its origin, giving rise to the field of jet substructure \cite{[{We are unable to do justice to the wealth of impressive experimental results using jet substructure. We refer the interested reader to \url{https://twiki.cern.ch/twiki/bin/view/AtlasPublic} and \url{http://cms-results.web.cern.ch/cms-results/public-results/publications/} for more examples}]results}. Of particular importance are multi-prong discriminants, which allow the identification of structures which are characteristic of hadronically decaying $W/Z/H$ bosons. Due to the complex environment of the LHC, these are used in conjunction with a grooming strategy, which removes low energy contamination. From a theoretical perspective, these techniques require understanding the structure of QCD jets at a much more differential level than previously considered. 

\begin{figure*}[t]
\subfigure{\raisebox{0.75cm}{(a)}\hspace{0.2cm}\includegraphics[width=6.4cm]{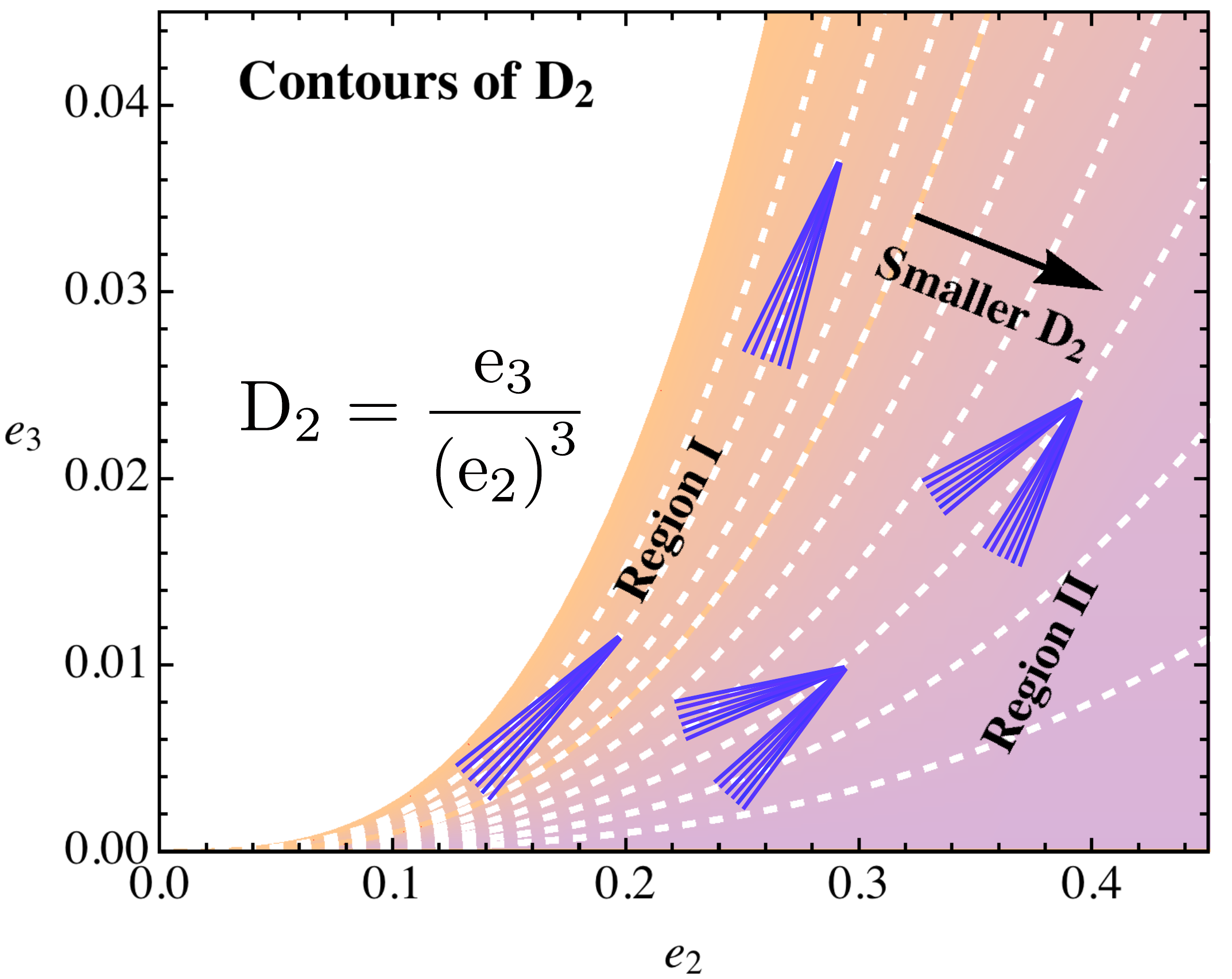}\label{fig:D2_ps}} \qquad \qquad \qquad
\subfigure{\raisebox{0.75cm}{(b)}\hspace{0.2cm}\includegraphics[width=6.2cm]{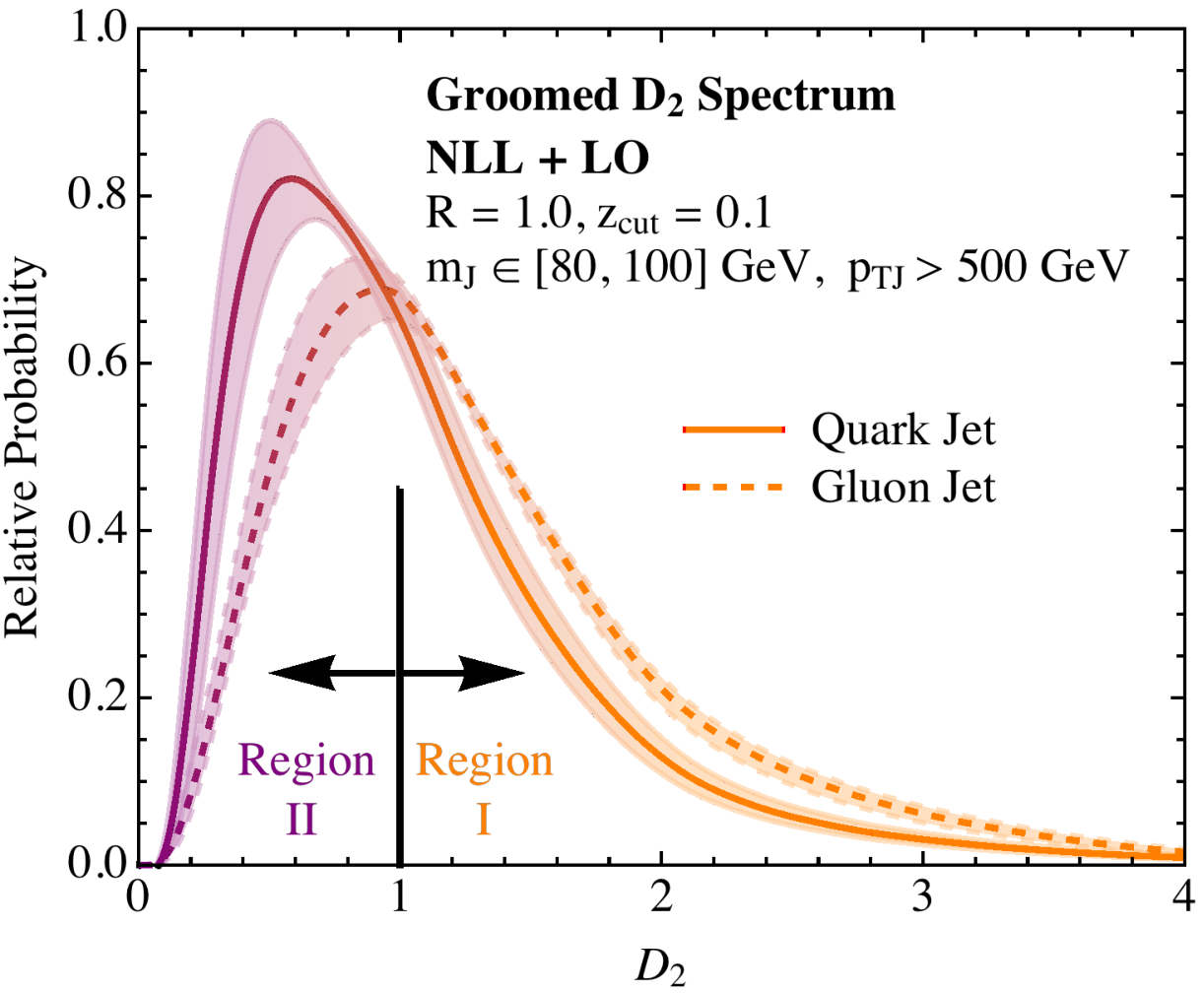}\label{fig:qg_dist}}
\caption{(a) Contours of $D_2$ in the $(e_2,e_3)$ phase space. (b) The groomed $D_2$ distribution obtained through marginalization of the multi-differential cross section. Effective field theories describing the different regions are discussed in the text.
}
\label{fig:multi_marginalize}
\end{figure*}

Analytic calculations play a crucial role in the field of jet substructure, having transformed it from relying on simple observables based on heuristics, to sophisticated observables which are able to exploit increasingly subtle aspects of gauge theories \cite{Larkoski:2013eya,Larkoski:2015npa,Maltoni:2016ays,Moult:2016cvt}, leading to improved performance and novel search strategies. As a concrete example, analytic calculations of previous status quo observables led to the modified mass drop (mMDT) \cite{Dasgupta:2013ihk,Dasgupta:2013via} and soft drop groomers \cite{Larkoski:2014wba}, as well as the $D_2$ \cite{Larkoski:2014gra,Larkoski:2015kga} and $N_2$ \cite{Moult:2016cvt} discriminants, which are the current tools of choice. Continued progress in developing new techniques relies on the next generation of calculations for deeper understanding and guidance.

In this paper we present a field theory framework for the analytic calculation of groomed multi-prong observables, building on recent developments in multi-scale effective field theories (EFTs) \cite{Bauer:2011uc,Larkoski:2015zka,Larkoski:2015kga,Pietrulewicz:2016nwo,Chien:2015cka}. It allows for a systematically improvable perturbative calculation based on operator definitions and a resummation of logarithmically enhanced terms, which dominate the behavior of the observable in the region of interest, using the renormalization group.

While our framework applies quite generally, in this paper we consider the specific case of the groomed $D_2$ observable \cite{Larkoski:2014gra,Larkoski:2015kga}, which has been widely used at the LHC to identify hadronically decaying $W/Z/H$ bosons. We compute for the first time at next-to-leading logarithmic (NLL) and leading order (LO) accuracy fully realistic distributions at the LHC, allowing us to draw robust theoretical conclusions about both the perturbative and non-perturbative behavior of the observable. 

\section{Jet Substructure Observables}
\label{sec:obs}

To identify hadronically decaying bosons based on their radiation patterns, we will use observables formed from the correlations of $p_T$ and boost invariant angle $R^2_{ij}=\phi_{ij}^2+\eta_{ij}^2$ \cite{Larkoski:2013eya,Moult:2016cvt}. These form the building blocks for observables of use at both ATLAS and CMS, and their simple structure facilitates analytic calculations and leads to well behaved perturbative corrections \cite{Larkoski:2015uaa}. Here we will focus on a specific example \cite{Larkoski:2014gra,Larkoski:2015kga}
\begin{align}\label{eq:D2_def}
D_2^{(\alpha)}= \frac{\ecf{3}{\alpha}}{(\ecf{2}{\alpha})^3}=\fd{1.3cm}{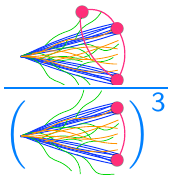}\,,
\end{align}
where the two-point and three-point correlation functions are defined as \cite{Larkoski:2013eya}
\begin{align}\label{eq:ppe2}
\ecf{2}{\alpha}&=\frac{1}{p_{TJ}^2}\sum_{i<j\in J} p_{Ti} p_{Tj}R_{ij}^\alpha\,, \\
\ecf{3}{\alpha}&=\frac{1}{p_{TJ}^3}\sum_{i<j<k\in J} p_{Ti} p_{Tj}p_{Tk}R_{ij}^\alpha R_{ik}^\alpha R_{jk}^\alpha\,.
\end{align}
In the remainder of the text, we will take $\alpha=2$, and drop the label to simplify notation. The $D_2$ observable measures the extent to which a jet has a two-prong substructure. For a two-prong jet, as is characteristic of a decaying $W/Z/H$, $D_2\ll1$, while for a more uniform jet, $D_2 \sim 1$, allowing them to be distinguished \cite{Larkoski:2014gra,Larkoski:2015kga}. This is illustrated in \fig{multi_marginalize}, which shows contours of constant $D_2$ in the $(e_2, e_3)$ phase space.

To reduce sensitivity to soft radiation, we will use the mMDT \cite{Dasgupta:2013ihk,Dasgupta:2013via} or soft drop \cite{Larkoski:2014wba} grooming algorithm. From a theoretical perspective, these algorithms have the advantage that they remove color connections to other jets \cite{Frye:2016okc,Frye:2016aiz}. After clustering a jet with the Cambridge/Aachen algorithm \cite{Dokshitzer:1997in,Wobisch:1998wt,Wobisch:2000dk}, the mMDT algorithm declusters the jet, and at each step compare branches $i$ and $j$, removing branches which fail the criteria 
\begin{equation}
\frac{\min[p_{Ti},p_{Tj}]}{p_{Ti}+p_{Tj}}>\zcut\,,
\end{equation}
where $\zcut$ is a parameter typically taken to be around $10\%$. The soft drop grooming algorithm generalizes this to include angular dependence, but in this paper we will only consider mMDT grooming for simplicity.

\section{Theoretical Framework}
\label{sec:theory}

To analytically compute substructure observables such as $D_2$ in a systematically improvable manner we will use techniques from effective theory, in particular, the soft-collinear effective theory (SCET) \cite{Bauer:2000yr,Bauer:2001ct,Bauer:2001yt,Bauer:2002nz,Rothstein:2016bsq}, which provides operator and Lagrangian techniques for analyzing factorization.

\subsection{Multi-Differential Marginalization}

Substructure discriminants of current interest take the form of ratios of IRC safe observables, such as \eq{D2_def}. These discriminants can be viewed as identifying contours (more generally hypersurfaces) in the multi-dimensional space spanned by the observables. The cross section can then be obtained by marginalization (integration along a contour) of the multi-differential cross section \cite{Larkoski:2013paa,Larkoski:2015kga}
\begin{align}
\frac{d\sigma}{dD_2}= \int d \ecfnobeta{2} d \ecfnobeta{3}\, \delta \left ( D_2 -\frac{ \ecfnobeta{3} }{( \ecfnobeta{2})^3 } \right ) \frac{d\sigma}{d \ecfnobeta{2} d \ecfnobeta{3}  }\,.
\end{align}
For the case of $D_2$, this is illustrated in \fig{multi_marginalize}, along with the final distributions for both quark and gluon jets at NLL+LO accuracy, as will be described in the text. 

The multi-differential cross section can be computed \cite{Larkoski:2014tva,Procura:2014cba} efficiently by tiling the multi-dimensional phase space with EFTs, which can then be smoothly patched together. This approach is completely general, and reduces the problem to identifying the correct EFTs in all asymptotic regions of the phase space \cite{Bauer:2011uc,Larkoski:2015zka,Larkoski:2015kga,Pietrulewicz:2016nwo}. The extension to additional measurements or subjets is conceptually simple by iterating these constructions.

\subsection{Factorization with Grooming}

In the case that $ m_J^2 \ll \zcut p_{TJ}^2\ll p_{TJ}^2$, which is satisfied for situations of interest at the LHC, the grooming procedure removes all wide angle soft radiation, including non-global logarithms \cite{Dasgupta:2001sh}, and the observable is determined by collinear physics \cite{Frye:2016okc,Frye:2016aiz}. In this limit there is also a well defined notion of quark and gluon jets \cite{Frye:2016okc,Frye:2016aiz}. Due to collinear factorization, the normalized cross section depends only on the quark and gluon fractions
\begin{align}
\frac{d\sigma^{ \text{norm}}}{d\Dobsnobeta{2}}=\sum\limits_{k} \kappa_k \frac{d\sigma_k^{\text{norm}}}{d\Dobsnobeta{2}}\,,
\end{align}
where the $\kappa_k$ can be interpreted as the fraction of jets with flavor $k$ within the given mass cut. It can be extracted using fixed order codes \cite{Campbell:1999ah,Campbell:2010ff,Campbell:2011bn,Alwall:2014hca}. We will drop the ``norm" superscript for simplicity

The grooming procedure, which isolates the collinear radiation from the rest of the event, makes the jet behave similar to a boosted event shape (other than the color flow), with center of mass energy set by the jet mass, $m_J$. For a color singlet decay this is strictly true, and was exploited in \cite{Feige:2012vc}. In the present case we will see that this will lead to a number of remarkable features, in particular for the structure of non-perturbative corrections. 

Although the dynamics of interest are purely collinear, energy hierarchies still exist for the radiation (modes) within the jet. 
Due to these hierarchies imposed by the $D_2$ measurement large logarithms exist in the perturbative calculation of the $D_2$ observable, which must be resummed to all orders to calculate the shape of the distribution. When used for discrimination, a cut $D_2 \ll 1$ is applied, making an accurate description of the distribution deep in the resummation region necessary.

We will perform the resummation by factorizing the cross section in each effective theory into single scale functions, $F$. Logarithms in the cross section are then resummed by the renormalization group evolution of these different functions 
$d\log F/d\log \mu =\gamma_{F}\,,$
where $\gamma_F$ is the anomalous dimension.

All functions have operator definitions that are valid to all orders using standard gauge invariant operators in SCET \cite{Bauer:2000yr,Bauer:2001ct}. These are either matching coefficients, which we denote $H$, jet functions, which describe energetic collinear dynamics, and are denoted $J$, or collinear-soft functions, which we denote $C_s$, capturing radiation with softer energies that couple eikonally to the energetic subjets. 

The jet functions are defined as matrix elements of collinear fields
\begin{align}
J_{n}(\ecfnobeta{3})&=\text{tr}\langle 0|\frac{\bar{n}\!\!\!\slash}{2}\chi_{n} \delta(Q-\bar{\mathcal{P}})\delta(\vec{{\mathcal P}}_{\perp})\delta(\ecfnobeta{3}-\ecfop{3})\bar{\chi}_{n}|0\rangle\,,
\end{align}
where $\chi_{n}(x) = \left[W_n^\dagger(x)\, \xi_n(x) \right]$, with $W_n$ a lightlike Wilson line, is a gauge invariant quark field, and $\cP$ is the momentum operator. Similar definitions exist for the gluon field. The collinear-soft functions are defined as matrix elements of Wilson line operators
\begin{align}\label{eq:cs_func}
\hspace{-1cm}C_{si}(\ecfnobeta{3})&=\text{tr}\langle 0|T\{Y_i\}\delta (\ecfnobeta{3}-\ecfop{3}) \Theta_{\text{SD}}\bar{T}\{Y_i\} |0\rangle\,.
\end{align}
Here $Y_i$ are products of Wilson lines along subjet, or the recoiling jet directions, $T$ and $\bar T$ denote time and anti-time ordering respectively, and $\ecfop{3}$ is an energy flow operator that implements the measurement function, while $\Theta_{\text{SD}}$ implements the soft drop constraints. These operators can be written in terms of the energy-momentum tensor of the effective theory \cite{Sveshnikov:1995vi,Korchemsky:1997sy,Lee:2006nr,Bauer:2008dt}.

As illustrated in \fig{multi_marginalize}, three distinct effective field theories are required to cover the entire phase space; an effective theory describing the region where no substructure is resolved within the jet (factorization $\text{I}$), and two effective theories describing when a two-prong substructure is resolved. The latter two are distinguished by the energy distribution of the subjets, namely if the subjets are both energetic, with $p_T\gg \zcut p_{TJ}$, which we will call $\text{II}_a$, or if one of the subjets has $p_T \sim \zcut p_{TJ}$, which we will call $\text{II}_b$. 

The factorization $\text{I}$ is a generalization of that given in \cite{Larkoski:2015kga}. Here the substructure of the jet is not resolved by the measurement, and the grooming acts only on dynamical soft radiation
\begin{align}\label{eq:fac_pp_cshaze}
\frac{d^2\sigma_k^{\text{I}}}{d\ecfnobeta{2} d\ecfnobeta{3}}= J(\ecfnobeta{2}) C_{s}(\ecfnobeta{2}, \ecfnobeta{3}, \zcut) \,.
\end{align}
The factorization $\text{II}_a$ is a generalization of \cite{Larkoski:2015kga}
\begin{align}\label{eq:fac_pp_coll}
\hspace{-0.5cm}\frac{d^2\sigma_k^{\text{IIa}}}{dz\, d\ecfnobeta{2} d\ecfnobeta{3}}= H^c(z,\ecfnobeta{2}) J_1(\ecfnobeta{3}) J_2(\ecfnobeta{3}) C_s(\ecfnobeta{3}, \zcut) \,,
\end{align}
Here $H^c$ is a matching coefficient which describes the splitting into two energetic subjets, $J_1$ and $J_2$, and is derived from collinear $1\to 2$ splitting amplitudes. The functions $C_s$ describe collinear-soft radiation at different energy scales, and $z$ is the $p_T$-fraction of one of the subjets of the total $p_{TJ}$.

The most interesting factorization occurs when the subjet energy, $p_T\sim \zcut p_{TJ}$. In this case, we can identify a universal matching coefficient that describes the splitting into a low energy jet in the presence of a grooming algorithm, analogous to the standard soft current \cite{Catani:2000pi,Duhr:2013msa,Li:2013lsa}. At leading power, $e_2$ is set by the Born-level kinematics of the subjets, making the result independent of the particular observable measured, and a universal ingredient in the study of groomed multiprong observables.
We have
\begin{align}\label{eq:fac_pp_cs}
\frac{d^2\sigma_k^{\text{IIb}}}{dz\,d\ecfnobeta{2} d\ecfnobeta{3}}=H^{s}(z,\ecfnobeta{2},\zcut )C_s(\ecfnobeta{3}) J_{sc}(\ecfnobeta{3}) J(\ecfnobeta{3}) \,,
\end{align}
In this case, the hard matching coefficient depends on $\zcut$. For example, the two diagrams that must be considered in the matching for the non-Abelian channel are 
\begin{align}\label{eq:hard_match}
\hspace{-0.3cm} H^{s}_{C_A}=\ \raisebox{0.4cm}{\fd{2.85cm}{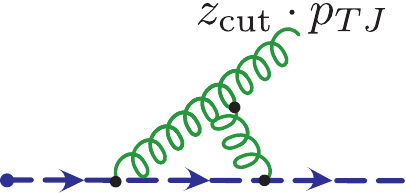}}\ +\ \raisebox{0.48cm}{\fd{3.15cm}{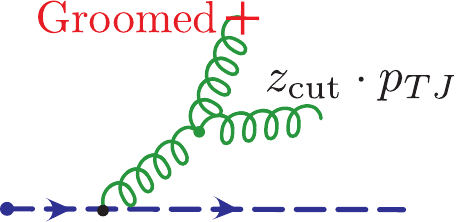}}
\end{align}
The first is a standard one-loop diagram, while in the second an additional real emission is removed by the mMDT algorithm. The one-loop anomalous dimension is 
\begin{align}
\gamma_{H^{s}} =&-\left(2C_F+C_A\right)\,\Gamma_{\text{cusp}}[\alpha_s]\log\frac{4\mu^2}{z\ecfnobeta{2}Q^2}-\frac{\alpha_s}{2\pi}\beta_0 \nn \\
&-\frac{\alpha_sC_F}{\pi}\,
\log\frac{z^2}{\zcut^2 }
+\frac{\alpha_s}{\pi}\left(
C_F-C_A
\right)\frac{\text{Cl}_2(\frac{\pi}{3})}{\pi} \,.
\end{align}
Here $\text{Cl}_2(\frac{\pi}{3})=\Im \text{Li}_2(e^{i\pi/3}) \approx 1.01494$ is the maximum value of the Clausen function. Polylogarithms of the sixth root of unity have appeared in a number of higher order calculations \cite{Broadhurst:1998rz,Kalmykov:2010xv,Bonciani:2015eua,Henn:2015sem}.
The leading logarithmic behavior is proportional to the cusp anomalous dimension, $\Gamma_{\text{cusp}}$ \cite{Korchemsky:1987wg}.

We have checked that summing over the different effective field theory contributions, and removing overlap \cite{Larkoski:2015kga,Pietrulewicz:2016nwo} our factorization correctly reproduces the small $D_2$ behavior by comparison with \eventtwo~\cite{Catani:1996vz}. 
 A more detailed discussion can be found in \cite{usD2l}.

\subsection{Non-Perturbative Corrections}

\begin{figure}[t]
\subfigure{\raisebox{2.55cm}{(a)}\hspace{0.2cm}\includegraphics[width=6.2cm]{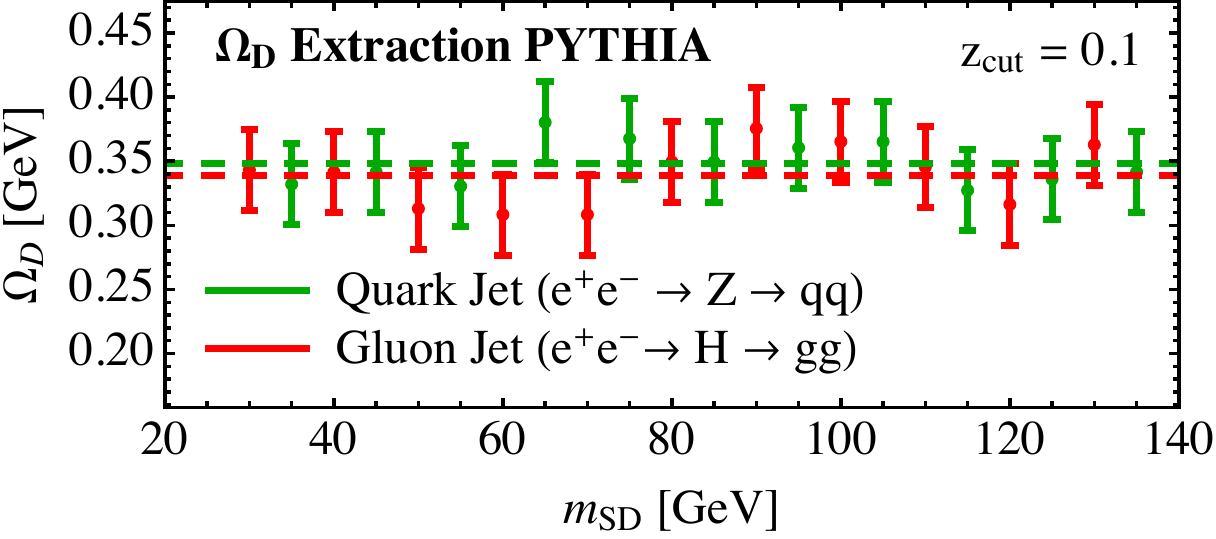}}\\
\subfigure{\raisebox{4.75cm}{(b)}\hspace{0.2cm}\includegraphics[width=6.2cm]{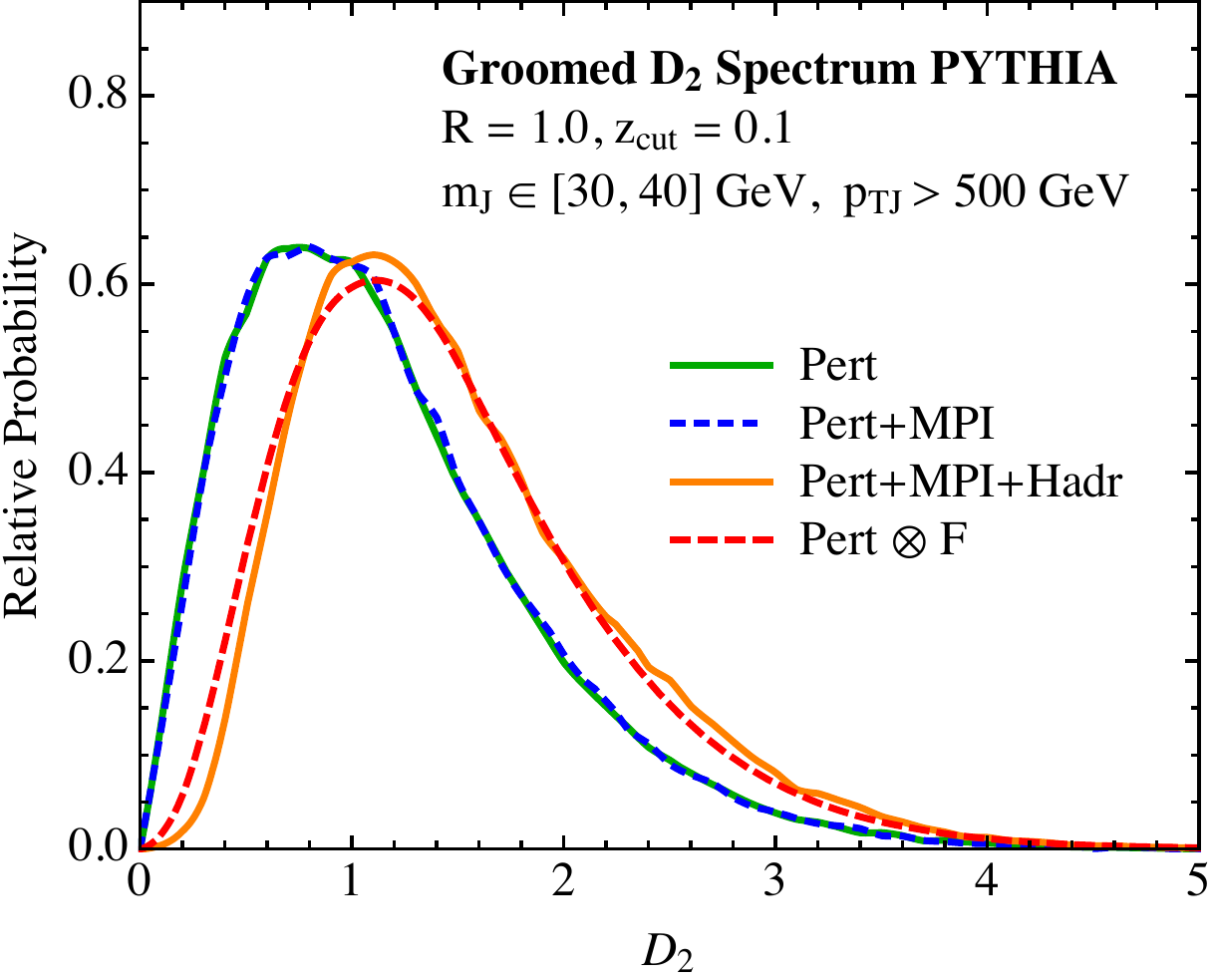}}
\caption{(a) Fit for the non-perturbative parameter $\Omega_D$ in $e^+e^-$ collisions. (b) Effect of MPI and hadronization as modeled by Pythia for $pp\to Zj$, and comparison with the shape function prediction.
}
\label{fig:np_shift}
\end{figure}

Non-perturbative corrections play an important role in the description of QCD event shapes. Our operator based formulation allows non-perturbative corrections to be defined in terms of matrix elements whose scalings and symmetry properties can be studied \cite{Korchemsky:1999kt,Korchemsky:2000kp,Lee:2006fn,Hoang:2007vb,Mateu:2012nk}. In the present case, the dominant non-perturbative corrections due to hadronization arise from collinear-soft modes, and can be shown to have a leading-power scaling as 
\begin{equation}\label{eq:npdef}
D_2^\text{NP}\sim \frac{\Omega_D}{\zcut^{3/2}m_J}\,.
\end{equation}
Here $\Omega_D$ is a non-perturbative scale, $\Omega_D\sim \Lambda_\text{QCD}$, and this expression is independent of the jet transverse momentum, $p_{TJ}$. The fact that the dominant non-perturbative effects arise from collinear-soft modes has two important consequences. First, the collinear-soft function, \eq{cs_func}, depends only on the color structure of the Wilson lines along the subjet and recoil directions, and its non-perturbative structure is therefore independent of the particular processes in which the jet was produced. Second, to leading order in the number of colors, the non-perturbative corrections are independent of whether a jet is gluon or quark initiated, since the dominant soft hadronization corrections arise from the zero rapidity region in the rest frame of a color-connected dipole, as was first experimentally established in \cite{Bartel:1983ii}.\footnote{This distribution of soft hadrons is manifest in the Lund string model \cite{Andersson:1983ia}, but applies to more general models of hadronization.} The outer dipoles will then have their contributions groomed away, but the interior will not. This is clearly seen from the color flow diagrams (here solid lines denote the color flows)
\begin{center}
\raisebox{-0.15cm}{\fd{2.65cm}{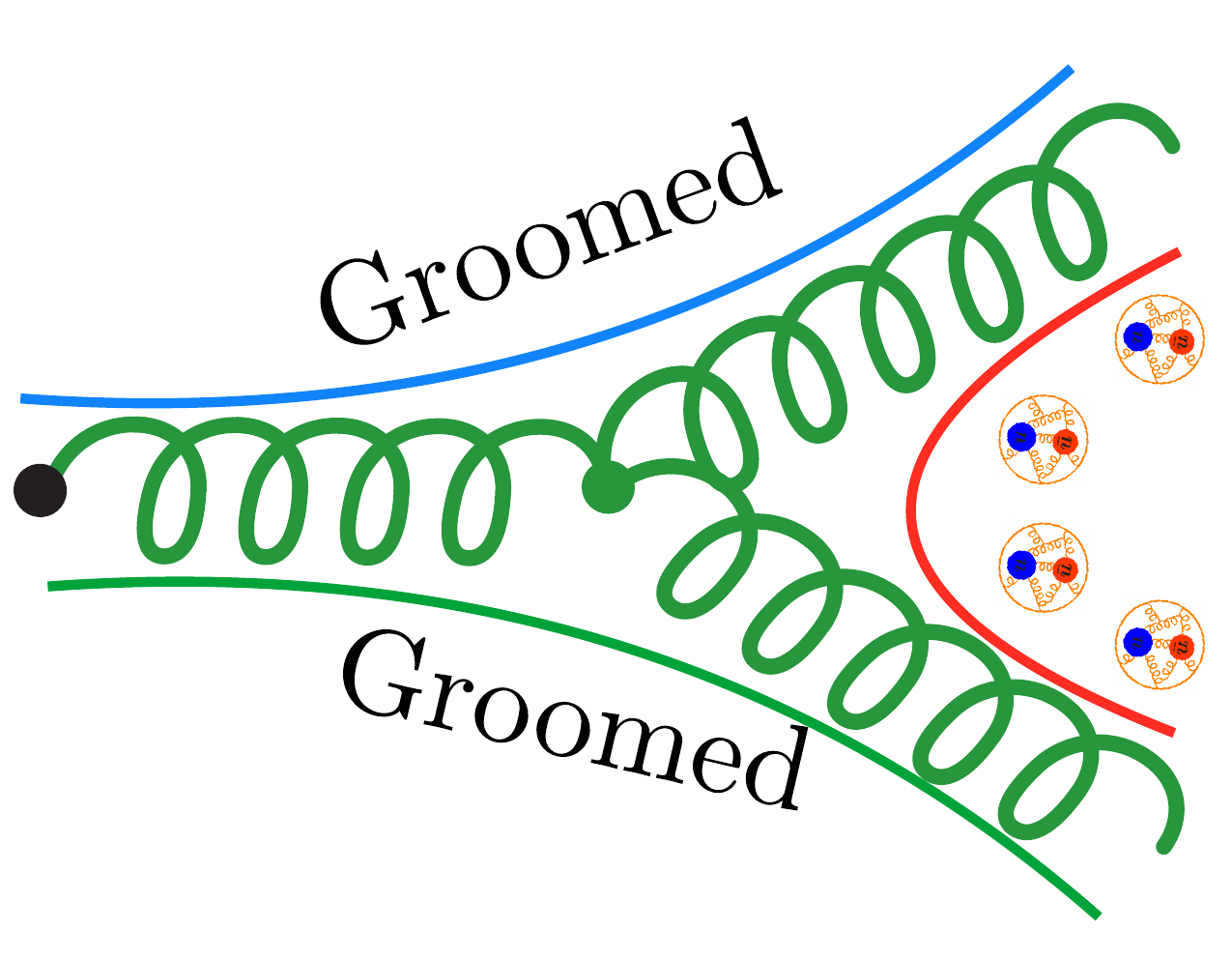}}\quad versus \quad \fd{2.65cm}{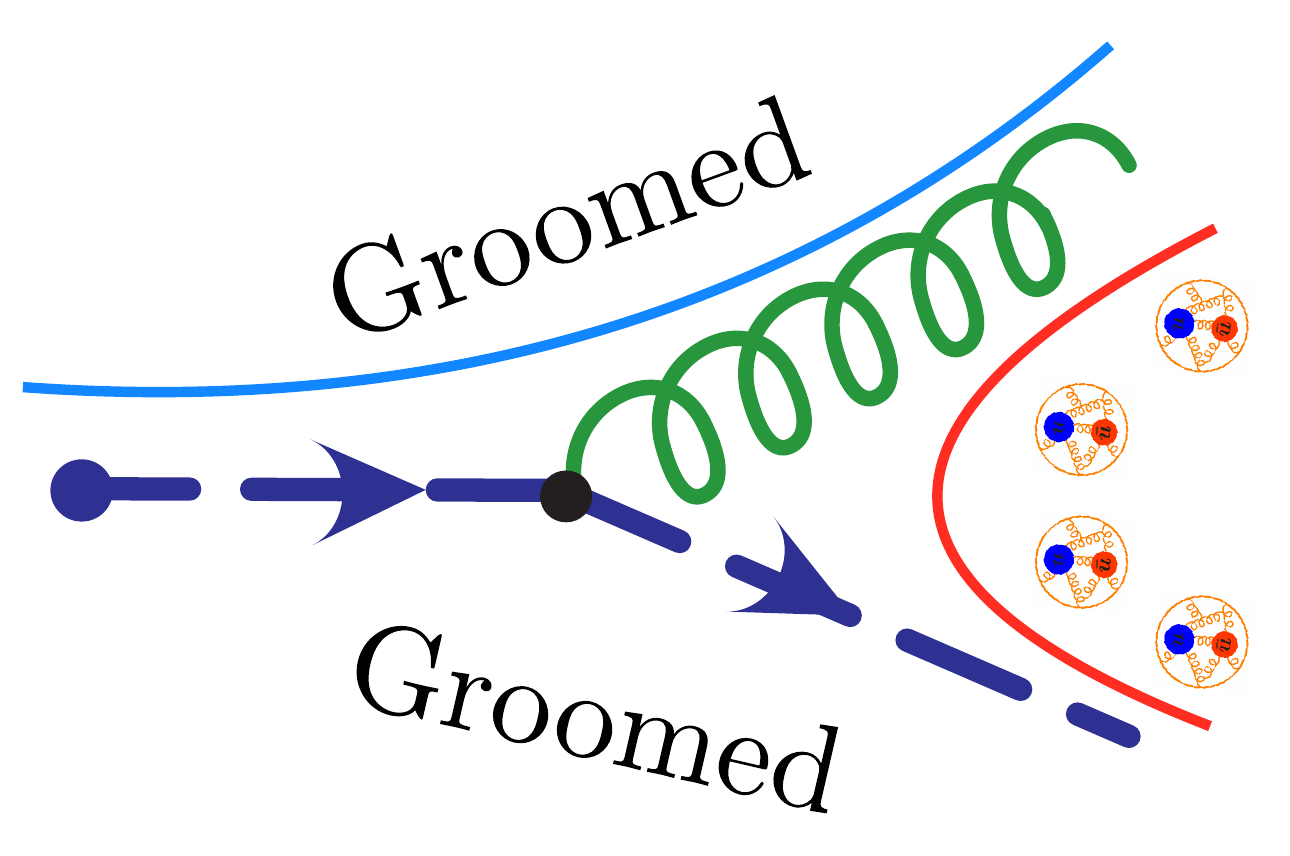}\,,
\end{center}
Contributions from the underlying event (which we model using multiple parton interactions (MPI)) have an identical scaling, but are further suppressed by the effective area of the jet, namely $m_J^2/p_{TJ}^2\ll \zcut$, and can therefore be completely neglected. We will verify that these predictions are well reproduced in simulation.

\begin{figure*}[t]
\subfigure{\raisebox{0.75cm}{(a)}\hspace{0.2cm}\includegraphics[width=6.95cm]{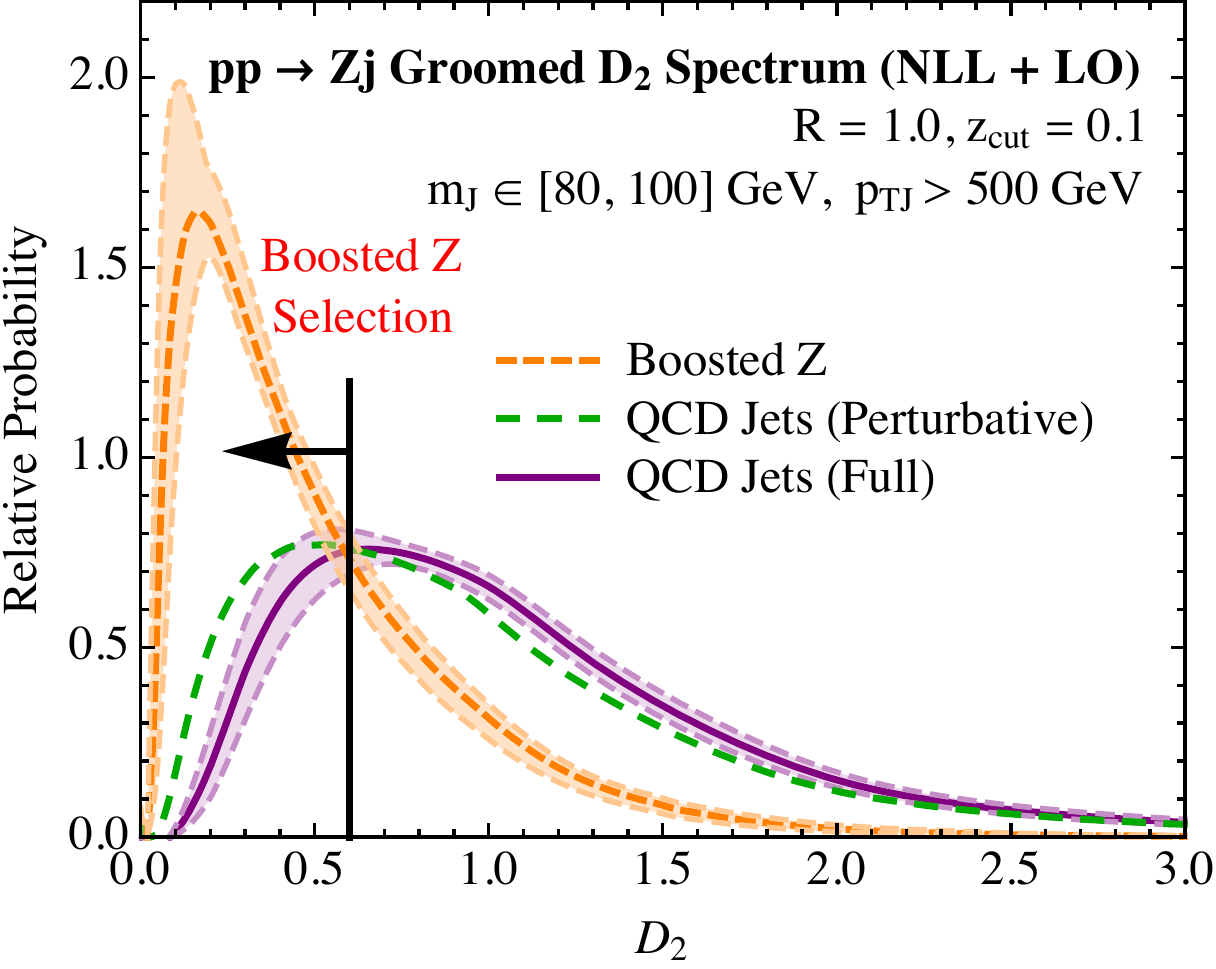} } 
\qquad \qquad
\subfigure{\raisebox{0.75cm}{(b)}\hspace{0.2cm}\includegraphics[width=7.3cm]{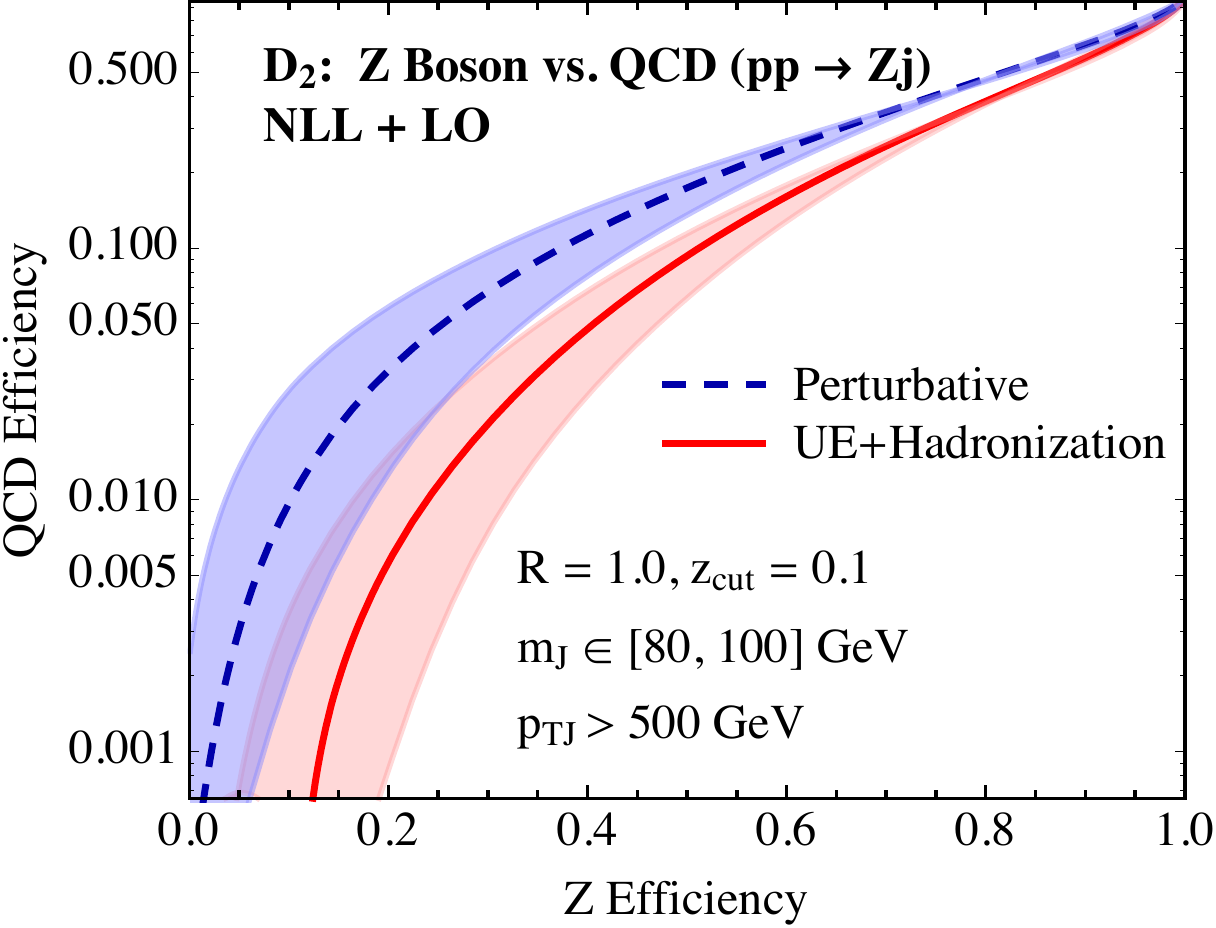} } 
\caption{(a) Analytic predictions for the groomed $D_2$ distribution for boosted $Z$ and QCD jets. (b) Prediction for the QCD efficiency vs.~$Z$ boson efficiency curve, illustrating the importance of non-perturbative corrections. 
}
\label{fig:Z}
\end{figure*}

Non-perturbative effects are implemented via a shape function $F(\epsilon)$, which is a parametrization of the non-perturbative matrix element, and are included via the convolution \cite{Korchemsky:1999kt,Korchemsky:2000kp}
\begin{align}
\frac{d\sigma_{\text{NP}}}{dD_2}=\int\limits_0^\infty d\epsilon\, F(\epsilon)   \frac{d\sigma}{dD_2} \left(  D_2 -\frac{\epsilon}{m_J \zcut^{3/2}} \right)\,.
\end{align}
 We take the functional form
$F(\epsilon)=(4\epsilon/\Omega_D^2) e^{-2\epsilon/\Omega_D}\,,$  \cite{Stewart:2014nna}
where $\Omega_D$ is defined as in \eq{npdef}.
By our factorization,  $\Omega_D$ is independent of the mechanism of jet production (up to possible flavor dependence, which we have argued is suppressed) and can therefore be estimated from \pythia{8.226} \cite{Sjostrand:2006za,Sjostrand:2014zea} in $e^+e^-$ collisions. As shown in \fig{np_shift}(a), we see that it is identical within errors for quarks and gluons and that it correctly reproduces the scaling with mass predicted by factorization. This implies that for the bulk of the $D_2$ distribution, only a single non-perturbative parameter is needed to describe groomed $D_2$ for all mass cuts and processes.

To verify this, in \fig{np_shift}(b) we test the effect of both MPI and hadronization for the $pp\to Zj$ process in \pythia{8.226}, and compare with the prediction of the shape function extracted from $e^+e^-$ collisions. We have chosen a smaller mass to enhance the non-perturbative effects. As predicted by factorization MPI is negligible, and the effect of hadronization is well reproduced by the shape function. We have verified this for a variety of other parameters and partonic channels, both using the perturbative factorization theorem prediction and the unhadronized \pythia{} distributions as input to the shape function, treating the parton shower with hadronization as data.

\section{Results for the LHC}
\label{sec:results}

Observables based on the energy correlation functions are being used in a wide range of applications at the LHC. Here we will illustrate our framework on a simple example, namely discriminating hadronically decaying $Z$-bosons from QCD. The extension to other processes is straightforward, using the factorization properties of the observable.

In \fig{Z}(a) we show analytic distributions for the $D_2$ observable for both a hadronically decaying $Z$ and QCD background at NLL matched to leading order fixed order computed using $1\to 3$ splitting functions. Quark and gluon fractions have been extracted after the application of a 80-100GeV mass cut. In \fig{Z}(b) we show an analytic signal efficiency vs.~background efficiency curve highlighting the difference between the perturbative and full results. While the non-perturbative effects have a moderate impact on the distribution, they have a non-trivial impact on the discrimination efficiency, particularly in the region of interest where $D_2$ is small, and therefore must be properly incorporated.

\section{Conclusions}
\label{sec:conc}

In this paper we have presented an effective field theory framework which allows for systematically improvable calculations for groomed multi-prong observables of current interest at the LHC. We demonstrated the power of our approach by computing for the first time the groomed $D_2$ observable at the LHC illustrating complete theoretical control over the perturbative and non-perturbative aspects of the observable. We hope that the next generation of theoretical understanding will drive the development of new jet substructure techniques to fully exploit the rich dataset of the LHC.

\section{Acknowledgements}

We thank Simone Marzani, Nhan Tran, Jesse Thaler, Phil Harris and Ben Nachman for helpful discussions and comments on the manuscript and Frank Tackmann for pictures of hadrons. 
This work is supported by the U.S. Department of Energy (DOE) under cooperative research agreements DE-FG02-05ER-41360, and DE-SC0011090 and by the Office of High Energy Physics of the U.S. DOE under Contract No. DE-AC02-05CH11231, and the LDRD Program of LBNL, as well as support from DOE contract DE-AC52-06NA25396 and through the LANL/LDRD Program.

\bibliography{sd_D2_bib}

\end{document}